\documentclass[reprint,superscriptaddress,amsmath,longbibliography,amssymb,prb]{revtex4-2}

\usepackage{graphicx}% Include figure files
\usepackage{dcolumn}% Align table columns on decimal point
\usepackage{bm}% bold math
\usepackage{multirow}
\usepackage{array}
\usepackage[usenames,dvipsnames]{xcolor}
\definecolor{nblue}{rgb}{0.0, 0.0, 1.0}
\definecolor{magenta}{rgb}{0.79, 0.08, 0.48}
\usepackage[colorlinks,linkcolor=nblue,urlcolor=blue,citecolor=blue,plainpages=false,pdfpagelabels,breaklinks]{hyperref}
\usepackage{float}

\newcommand{\beq}{\begin{equation}}
\newcommand{\eeq}{\end{equation}}
\newcommand{\bea}{\begin{eqnarray}}
\newcommand{\eea}{\end{eqnarray}}

\begin{document}
%--------------------------------------------------------------title----------------------------------------------------
\title{Prediction of high-temperature ambient-pressure superconductivity in hexagonal boron-rich clathrates }

%--------------------------------------------------------------author---------------------------------------------------

\author{Bin Li}\email[Electronic addresses: ]{libin@njupt.edu.cn}
\affiliation{School of Science, Nanjing University of Posts and Telecommunications, Nanjing 210023, China}
\affiliation{Jiangsu Provincial Engineering Research Center of Low Dimensional Physics and New Energy, Nanjing University of Posts and Telecommunications, Nanjing 210023, China}

\author{Yuxiang Fan}
\affiliation{School of Science, Nanjing University of Posts and Telecommunications, Nanjing 210023, China}

\author{Chuanhui Yin}
\affiliation{College of Electronic and Optical Engineering, Nanjing University of Posts and Telecommunications, Nanjing 210023, China}

\author{Junjie Zhai}
\affiliation{School of Science, Nanjing University of Posts and Telecommunications, Nanjing 210023, China}
\affiliation{Key Laboratory of Dark Matter and Space Astronomy, Purple Mountain Observatory, Chinese Academy of Sciences, Nanjing 210023, China}

\author{Cong Zhu}
\affiliation{College of Electronic and Optical Engineering, Nanjing University of Posts and Telecommunications, Nanjing 210023, China}

\author{Zhisi Cao}
\affiliation{School of Science, Nanjing University of Posts and Telecommunications, Nanjing 210023, China}
%\affiliation{University of Chinese Academy of Sciences Nanjing (UCASNJ), Nanjing 211135, China}

\author{Jie Cheng}
\affiliation{School of Science, Nanjing University of Posts and Telecommunications, Nanjing 210023, China}
\affiliation{Jiangsu Provincial Engineering Research Center of Low Dimensional Physics and New Energy, Nanjing University of Posts and Telecommunications, Nanjing 210023, China}

\author{Shengli Liu}
\affiliation{School of Science, Nanjing University of Posts and Telecommunications, Nanjing 210023, China}

\date{\today}

%------------------------------------------------------Abstract---------------------------------------------------

\begin{abstract}
Inspired by recent predictions of superconductivity in B-C framework clathrates, we employ density functional theory to explore potential superconductors among hexagonal hydride-substituted compounds with compositions XB$_8$C, XB$_7$C$_2$, XB$_6$C$_3$, XB$_3$C$_6$, XB$_2$C$_7$, and XBC$_8$. Our high-throughput calculations on 96 compounds reveal several dynamically stable candidates exhibiting superconductivity at ambient pressure. Analysis of electronic structures and electron-phonon coupling demonstrates that CaB$_8$C, SrB$_8$C, and BaB$_8$C possess superconducting transition temperatures ($T_c$) exceeding 50 K, with CaB$_8$C exhibiting the highest predicted $T_c$ of 77.1 K among all stable compounds studied. These findings expand the family of B-C clathrate superconductors and provide valuable insights for experimental efforts aimed at discovering novel superconducting materials.
\end{abstract}
%\noindent{\bf Keywords:}
%\noindent{\it Superconductivity, B-C clathrates, High-throughput calculations, Ambient pressure\/}\\
\maketitle

%-----------------------------------------------------Introduction------------------------------------------------
\section{Introduction}
The concept of ``metallic hydrogen'' first proposed in 1935 \cite{wigner1935possibility}, was later expanded by Ashcroft, who predicted its potential for room-temperature superconductivity under high pressure \cite{ashcroft1968metallic}. Ashcroft further suggested that hydrogen-rich materials could achieve high-temperature superconductivity at lower pressures through ``chemical pre-compression'' \cite{ashcroft2004hydrogen}. Recent theoretical studies have indicated that hydrogen-based superconductors can exceed room-temperature transition temperatures, with many hydrogen-rich compounds containing main group elements now considered potential high-$T_c$ superconductors. Experimental verifications include H$_3$S (203 K at 200 GPa) \cite{drozdov2015conventional}, MgH$_6$ (271 K at 300 GPa) \cite{feng2015compressed}, CaH$_6$ (235 K at 150 GPa) \cite{wang2012superconductive}, LaH$_{10}$ (280 K at 210 GPa) \cite{drozdov2019superconductivity}, and CeH$_9$ (105-117 K at 200 GPa) \cite{salke2019synthesis}. However, the extreme pressure conditions exceeding 100 GPa present significant challenges for practical applications \cite{ma2022high, geballe2018synthesis}. To address this limitation, several ternary hydrides have been reported with more favorable superconducting properties at lower pressures, including LaBeH$_8$ (110 K at 80 GPa) \cite{song2023stoichiometric}, LaBH$_8$ (126 K at 50 GPa) \cite{di2021bh}, CeBeH$_8$ (56 K at 10 GPa), CeBH$_8$ (118 K at 150 GPa) \cite{hou2022superconductivity}, \textcolor[rgb]{0.00,0.00,0.00}{ and (La,Y)H$_{10}$ (253 K at 183 GPa) \cite{semenok2021superconductivity}} opening new avenues for high-temperature superconductor research and potential applications.

The stability challenges of hydride-rich high-temperature superconductors at low pressures have shifted scientific focus towards boron-carbon/boron-nitride\textcolor[rgb]{0.00,0.00,0.00}{/borosilicide}  superconductors \cite{wang2021high, geng2023conventional, cui2022prediction, cui2020rbb3si3, hai2022improving, gai2022van, zhang2023machine, li2024superconductivity}, which exhibit stability at low or ambient pressures. Carbon, a fundamental element, forms diverse stable structures with other elements and itself, as evidenced by superconductivity in graphite intercalation compounds \cite{iye1982superconductivity}, CaC$_6$ ($T_c = 11.5$ K) \cite{emery2005superconductivity}, and Cs$_3$C$_{60}$ and RbCs$_2$C$_{60}$ ($T_c = 40$ K and 33 K, respectively) \cite{palstra1995superconductivity, tanigaki1991superconductivity}. Boron, with its unique electron-deficient configuration, contributes to versatile bonding in metal borides like MgB$_2$, a conventional superconductor with $T_c = 39$ K at atmospheric pressure due to strong electron-phonon coupling \cite{nagamatsu2001superconductivity}. The investigation of boron-carbon bonding networks, including boron-doped diamond \cite{ekimov2004superconductivity}, SrB$_3$C$_3$ \cite{zhu2023superconductivity}, Ca-B-C systems \cite{zhang2023machine}, KPbB$_6$C$_6$, and Rb$_{0.4}$Sr$_{0.6}$B$_3$C$_3$ \cite{geng2023conventional, gai2022van}, has been motivated by the concept of substituting carbon atoms with boron to stabilize structures \cite{zeng2015li}. This approach offers a promising avenue for designing superconductors with unique properties, as the prevalence of sp$^3$ hybridization in these compounds enables strong electron-phonon coupling, enhancing superconductivity while reducing pressure requirements for dynamic stability. The exploration of boron-carbon and boron-nitrogen frameworks, driven by their inherent structural stability and potential for low-pressure high-$T_c$ superconductivity, represents a promising frontier in condensed matter physics and materials science.

Building upon previous research \cite{10.1063/1.5130583,salke2019synthesis}, we propose \textcolor[rgb]{0.00,0.00,0.00}{an approach} to superconductivity by substituting hydrogen with boron and carbon elements, aiming to achieve high-temperature superconductivity at lower pressures. Our theoretical study focuses on the stoichiometry of B and C, emphasizing compound stability under atmospheric conditions and the potential for high-$T_c$ superconductivity. Using first-principles calculations, we investigate the electronic band structure, phonon spectrum, electron-phonon interaction, and $T_c$ of compounds with the general formula XB$_8$C at ambient pressure, where $X$ represents Ca, Ba, or Sr. Notably, our findings predict a $T_c$ of 77.1 K for CaB$_8$C at ambient pressure, significantly expanding the potential of B-C framework inclusion complexes in the field of superconductivity. This work not only explores a new class of superconducting materials but also provides insights into the fundamental mechanisms underlying high-temperature superconductivity in non-hydride systems, potentially paving the way for the design of practical, ambient-pressure superconductors.

%------------------------------------------------------Methods------------------------------------------------------
\section{Methods}
Electronic bands, density of states, and Fermi surfaces were calculated using the WIEN2K package, which implements the full potential linear augmented plane wave (FP-LAPW) method with local orbitals \cite{blaha2001wien2k, blaha1990full}. The Pedrew-Burke-Ernzerhof form of the generalized gradient approximation (GGA) was chosen for the exchange-correlation functional, providing values closer to experimental results. Phonon and electron-phonon coupling matrix elements were computed using density functional perturbation theory (DFPT) \cite{giannozzi2020quantum}, with pseudopotentials selected from the standard solid-state pseudopotential (SSSP) library \cite{prandini2018precision}. The Quantum-Espresso (QE) package, utilizing the pseudopotential plane wave method, was employed within the DFPT framework to obtain detailed structural information \cite{giannozzi2009quantum}. Self-consistent calculations used a plane wave basis set with charge density and wave function cutoffs of 600 Ry and 60 Ry, respectively. Convergence was achieved using a 6$\times$6$\times$6 high-symmetry $q$-point grid and a 12$\times$12$\times$12 $k$-point grid for crystal structure optimization and total energy estimation. Efficient materials calculations employed dense 24$\times$24$\times$24 Monkhorst-Pack special points \cite{monkhorst1976special}, while Brillouin zone (BZ) integration utilized the tetrahedron method \cite{kawamura2014improved}. Crystal structure modeling and visualization were performed using VESTA \cite{momma2011vesta}, with Fermi surface analysis conducted using XCRYSDEN and FermiSurfer tools \cite{kawamura2019fermisurfer}. To characterize the superconducting properties of the most promising candidate materials, we employed the Eliashberg spectral function $\alpha^2F(\omega)$, which effectively represents the electron-phonon coupling in conventional superconductivity:
\begin{equation}
{\alpha ^2}{\rm{F(}}\omega {\rm{) = }}\frac{1}{{2\pi N(E_{f})}}\sum\limits_{Q\nu}^{} {\frac{{\gamma _{Q\nu}}}{{\hbar\omega _{Q\nu}}}\delta (\omega  - \omega _{Q\nu})},
\end{equation}

where $N$($E_{f}$) is density of states at the Fermi level, $\gamma_{Q\nu}$ is the electron-phonon line width, and $\omega_{Q\nu}$ is the phonon frequency at phonon branch $\nu$ and wavevector $Q$. The electron-phonon coupling constant $\lambda$ is obtained by integrating $\alpha$$^2$$F$($\omega$) and is written as: 

\begin{equation}
\lambda=\sum\limits_{Q\nu}^{} {\lambda_{Q\nu}}= 2\int_{0}^{\omega}d\omega\alpha^2F(\omega)/\omega,
\end{equation}

where $\omega$$_l$$_o$$_g$ is the logarithmic average phonon frequency, expressed as:
\begin{equation}
{\omega _{\log }} = \exp [\frac{2}{\lambda }\int_0^\infty  {\frac{{d\omega }}{\omega }{\alpha ^2}{\rm{F}}(\omega )\log(\omega) } ].
\end{equation}

The approximate estimation of the transition temperature is obtained using the Allen-Dynes modified McMillan (ADM) formula. Typically, the predicted critical temperature closely aligns with the critical temperature derived from the solution of the Eliashberg equations \cite{allen1975transition}. The ADM  equation is as follows:

\begin{equation}
{T_c} = f_1f_2\frac{{{\omega _{log}}}}{{1.2}}{\rm{exp}}\left[ { - \frac{{1.04(1 + \lambda )}}{{\lambda  - {\mu ^*}(1 + 0.62\lambda )}}} \right],
\end{equation}

The Coulomb pseudopotential constant $\mu^*$ is set to 0.1 in these calculations. The factor $f_{1}$ accounts for strong coupling corrections, while the factor $f_{2}$ represents shape correction. The values of these factors, $f_{1}$ and $f_{2}$, are determined by the parameters $\lambda$, $\mu^*$, $\omega_{\text{log}}$, and $\overline{\omega_2}$. The expressions for these factors are given by:

\begin{equation}
f_{1} = \sqrt[3]{1 + \left(\frac{\lambda}{2.46(1 + 3.8\mu^*)}\right)^{\frac{3}{2}}}
\end{equation}

\begin{equation}
f_{2}= 1+\frac{(\overline{\omega _2}/\omega_{log}-1)\lambda^2}{\lambda^2+3.312(1+6.3\mu^*)^2(\overline{\omega _2}/\omega_{log})^2}
\end{equation}

%---------------------------------------------Results and discussion-------------------------------------------------
\section{Results and discussion }

\begin{figure}
	\begin{center}
		\includegraphics[width=9cm]{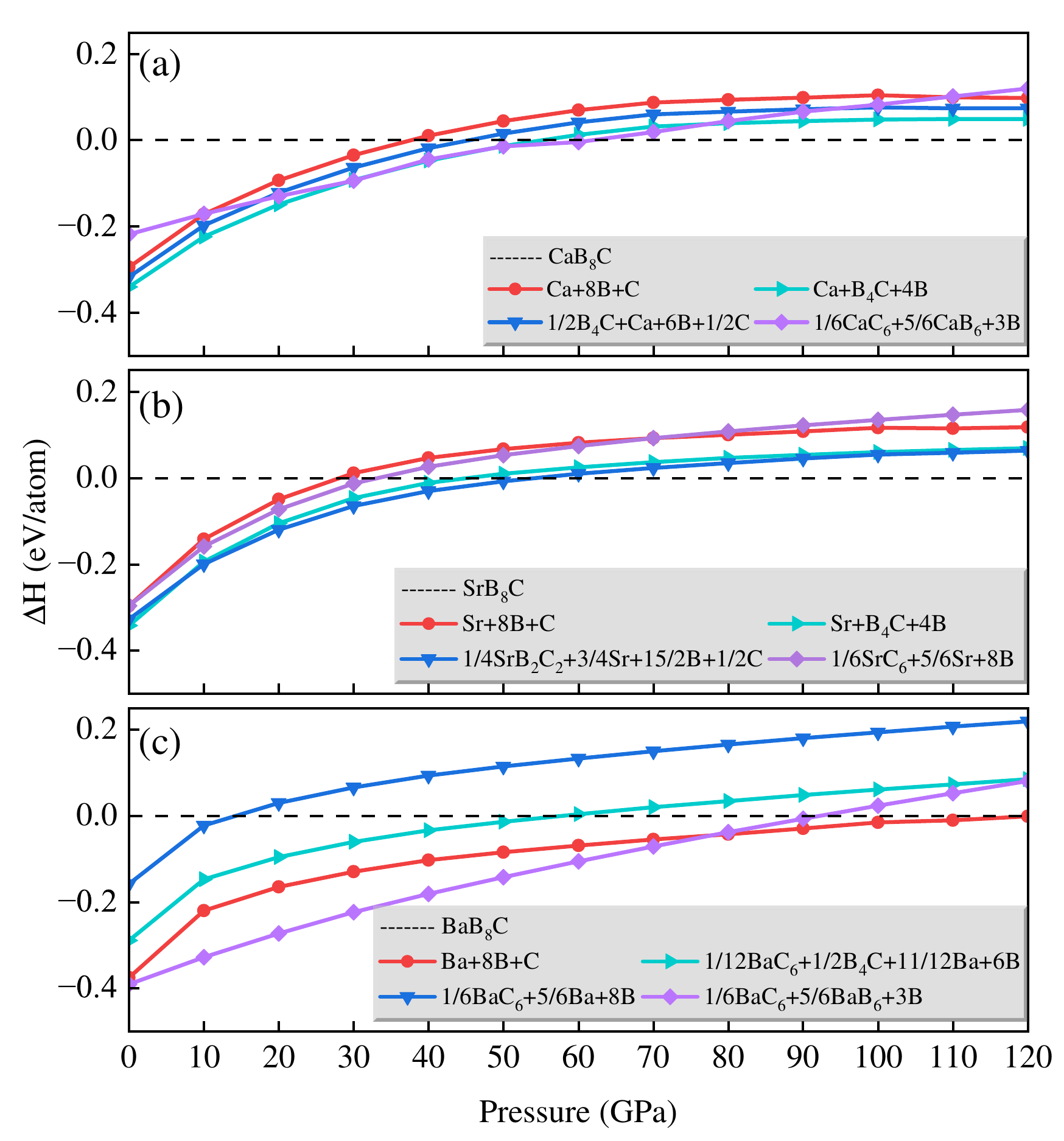}
		\caption{The relationship between the CaB$_8$C, SrB$_8$C and BaB$_8$C structure and the enthalpy difference of the different products of decomposition is calculated for pressures ranging from 0 to 120 GPa.}
		\label{fig1}
	\end{center}
\end{figure}
%We examined its decomposition enthalpy across a pressure range of 0 to 160 GPa, revealing that CaB$_8$C becomes thermodynamically stable above 40 GPa, transitioning from its decomposed state of Ca + 8B + C. Based on calculated kinetic stability and electron-phonon coupling, XB$_8$C ($X$ = Ca, Sr, Ba) emerged as the primary candidates for further investigation. In the main script, our detailed analysis focused specifically on XB$_8$C due to their superior electron-phonon coupling and superconducting properties.
Prior to analyzing the stability of ternary compounds in the boron-carbon system at ambient pressure, we evaluated their structural properties by substituting hydrogen atoms in CeH$_9$ with boron and carbon atoms, using the known $P6_3/mmc$ (space group No. 194) structure as a reference~\cite{salke2019synthesis}. 
%This substitution results in six potential compound configurations: B$_8$C, B$_7$C$_2$, B$_6$C$_3$, B$_3$C$_6$, B$_2$C$_7$, and BC$_8$, in which metal atoms are replaced by Mg, Li, Be, Al, Ba, Lu, La, Na, K, Y, Rb, Sc, Sr, Cs, Ca, or Ce. The compounds' characteristics differ significantly depending on the chemical ratios of B and C atoms. We finally settled on the XB$_8$C structure after high-throughput computations on all conceivable compounds that can retain kinetic stability while achieving superconductivity., in which alkali metal atoms occupy cage locations, ~\cite{wood2011intercalated, saal2013materials, mccarty1958new, pauling1934kurzere, shamp2017properties, ott1997structure, schmitt2001crystal} SrB$_6$ ~\cite{ott1997structure},B~\cite{mccarty1958new},We finally settled on the XB$_8$C structures after high-throughput computations on all conceivable compounds that can retain kinetic stability while achieving superconductivity. We next ran electron-phonon coupling calculations on the class of compounds, and the results narrowed our attention to Ca, Sr, and Ba.
This substitution results in six compound configurations: XB$_8$C, XB$_7$C$_2$, XB$_6$C$_3$, XB$_3$C$_6$, XB$_2$C$_7$, and XBC$_8$. We evaluated  16 potential $X$ atoms ($X$ = Li, Na, K, Rb, Cs, Be, Mg, Ca, Sr, Ba, Al, Sc, Y, La, Lu, and Ce). The characteristics of these compounds exhibit substantial variation based on the precise chemical composition of boron and carbon atoms. Through extensive high-throughput computational screening, we systematically explored all potentially stable compounds. Subsequent electron-phonon coupling calculations further narrowed our focus to the elements Ca, Sr, and Ba, as their XB$_8$C compounds exhibit electron-phonon coupling constants greater than 1. We examined the decomposition enthalpies of XB$_8$C ($X$ = Ca, Sr, Ba) in the pressure range of 0 to 120 GPa and found that above 40 GPa, CaB$_8$C transitions from the decomposition state of Ca + 8B + C, and SrB$_8$C transitions from the decomposition state of Sr + B$_4$C + 4B. Several additional synthesis routes were identified: \textcolor[rgb]{0.00,0.00,0.00}{$\frac{1}{2}$B$_4$C + Ca + 6B + $\frac{1}{2}$C $\stackrel{45 \ \text{GPa}}{\longrightarrow}$ CaB$_8$C, $\frac{1}{6}$CaC$_6$ + $\frac{5}{6}$CaB$_6$ + 3B $\stackrel{60 \ \text{GPa}}{\longrightarrow}$ CaB$_8$C,
Sr + 8B + C $\stackrel{30 \ \text{GPa}}{\longrightarrow}$ SrB$_8$C, 
$\frac{1}{6}$SrC$_6$ + $\frac{5}{6}$Sr + 8B $\stackrel{35 \ \text{GPa}}{\longrightarrow}$ SrB$_8$C, 
$\frac{1}{4}$SrB$_2$C$_2$ + $\frac{3}{4}$Sr + $\frac{15}{2}$B + $\frac{1}{2}$C $\stackrel{55 \ \text{GPa}}{\longrightarrow}$ SrB$_8$C,
$\frac{1}{6}$BaC$_6$ + $\frac{5}{6}$Ba + 8B $\stackrel{13 \ \text{GPa}}{\longrightarrow}$ BaB$_8$C,
Ba + 8B + C $\stackrel{120 \ \text{GPa}}{\longrightarrow}$ BaB$_8$C,}
as illustrated in Fig.~\ref{fig1}. The decomposition products were primarily identified from structures examined in the Open Quantum Materials Database (OQMD)~\cite{saal2013materials}, including CaC$_6$~\cite{wood2011intercalated}, CaB$_6$~\cite{pauling1934kurzere}, B$_4$C~\cite{shamp2017properties}, and BaB$_6$~\cite{schmitt2001crystal}. This comprehensive analysis of structural stability and pressure-dependent synthesis routes provides crucial insights into the behavior of CaB$_8$C\textcolor[rgb]{0.00,0.00,0.00}{, SrB$_8$C and BaB$_8$C, }laying the foundation for subsequent investigations of its potential superconducting properties.

%The pressure required for BaB$_8$C to reach a stable state after decomposition is higher.

\begin{figure}
\begin{center}
\includegraphics[width=8cm]{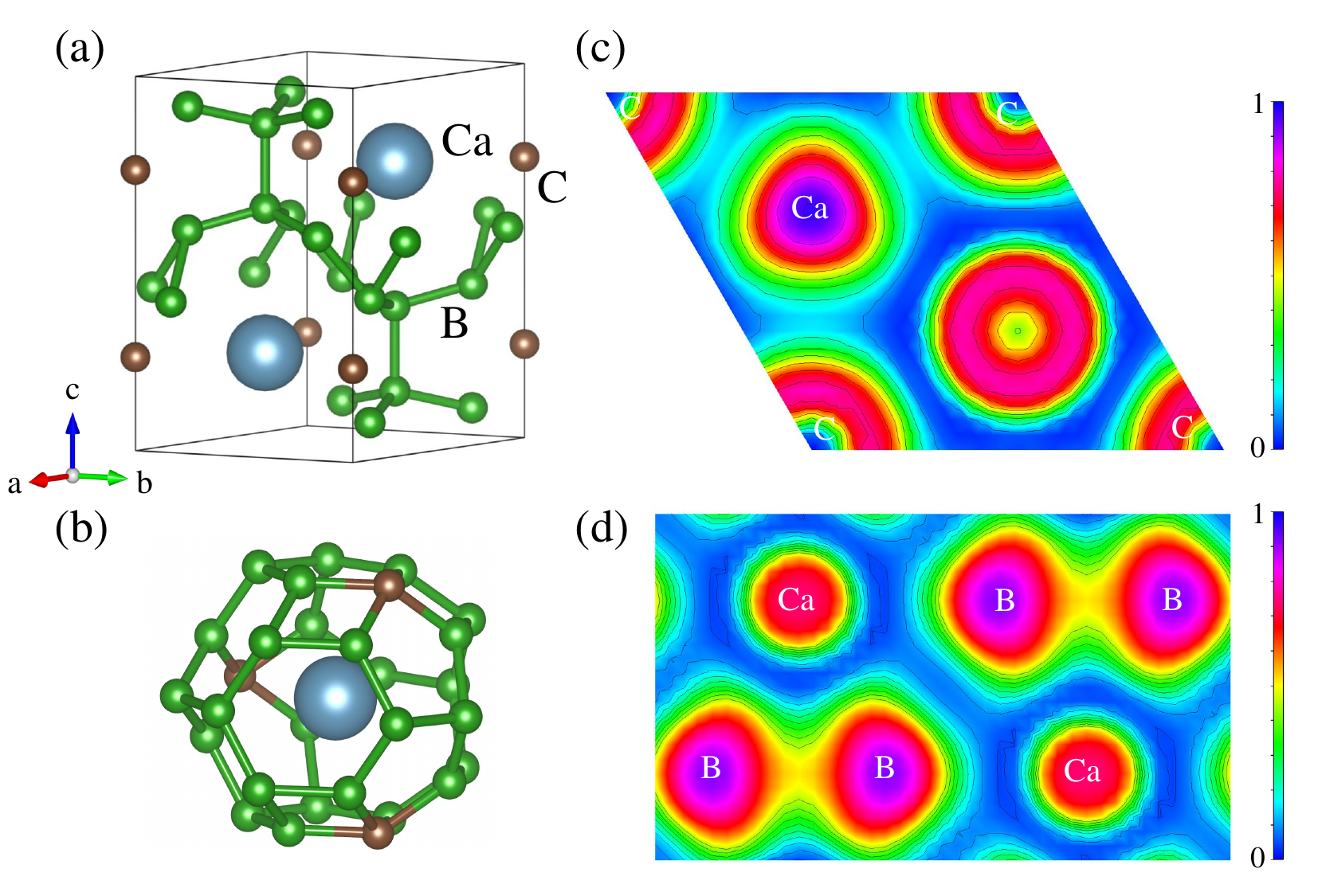}
\caption{Structure and electron localization in $P6_3/mmc$-CaB$_8$C. (a) Crystal structure of the hexagonal $P6_3/mmc$ phase of CaB$_8$C. (b) Boron-carbon cage encapsulating a central calcium atom. (c) Calculated ELF in the (001) plane and (d) ELF in the (100) plane.}

\label{fig2}
\end{center}
    \end{figure}

Figure~\ref{fig2}(a) depicts the optimized hexagonal structure of the CaB$_8$C unit cell in the $P6_3/mmc$ space group at ambient pressure. Unlike most boron-carbon superconductors, the XB$_8$C compound studied here exhibits a relatively high concentration of B atoms. In this structure, Ca atoms occupy the Wyckoff position 2$d$ (1/3, 2/3, 1/4), C atoms are located at 2$b$ (0, 0, 1/4), and B atoms are distributed at 4$f$ (1/3, 2/3, 0.1499) and 12$k$ (0.1565, 0.8435, 0.4404). Figure ~\ref{fig2}(b) illustrates the boron-carbon cage that encapsulates a central calcium atom. The optimized CaB$_8$C has lattice constants a = b = {4.679} Å and c = {7.7596} Å. 
%Structural optimization of CaB$_8$C, SrB$_8$C, and BaB$_8$C reveals a monotonic increase in the B-C bond length: from {1.684} Å in CaB$_8$C to {1.706} Å in SrB$_8$C, and further to {1.728} Å in BaB$_8$C. This trend may be attributed to variations in charge distribution and transfer within the covalent bonds, as well as complex Coulomb interactions among the atoms. The electron localization function (ELF) along the (001) plane, shown in Fig.~\ref{fig2}(c), indicates that Ca contributes a significant amount of charge, while the C atom provides few electrons. In the (100) plane [Fig.~\ref{fig2}(d)], strong covalent bonds are observed between B atoms, likely contributing to the stability of CaB$_8$C under ambient pressure.
Structural optimization of CaB$_8$C, SrB$_8$C, and BaB$_8$C reveals a monotonic increase in the B--C bond length. \textcolor[rgb]{0.00,0.00,0.00}{As the atomic mass and ionic radius of the alkaline earth metal increase along the series Ca $\rightarrow$ Sr $\rightarrow$ Ba, the B--C bond length systematically increases from {1.684} Å in CaB$_{8}$C to {1.706} Å in SrB$_{8}$C and further to {1.728} Å in BaB$_{8}$C. The electron localization function (ELF) in Figure~\ref{fig2}(c,d) illustrates the covalent B--B bonding. Using Bader charge analysis, we quantified the electron transfer for each element in the structure. The analysis reveals that each Ca atom donates 1.409 electrons (charge state: +1.409), contributing a total of 2.818 electrons to the B--C cage. Within the boron framework, four B atoms act as electron acceptors while twelve B atoms serve as donors. Each C atom gains an average of 3.658 electrons, resulting in a formal charge of -3.658. Detailed Bader charge analysis for each element can be found in Table S1 of the supplementary material\cite{SupplementalMaterial}.}

%{As the atomic mass and ionic radius of the alkali metal increase, the B-C bond length increases from {1.684} Å in CaB$_8$C to {1.706} Å in SrB$_8$C and further to {1.728} Å in BaB$_8$C. Fig.~\ref{fig2}(c) and (d) clearly illustrate the covalent bonding between B-B, which may contribute to the stability of CaB$_8$C at ambient pressure. There are localized electrons between the B and C atoms, along with appropriate electron distributions around the Ca and B atoms. We employed the Bader method to calculate the electron transfer for each element in the structure. The Bader analysis indicates that each Ca atom loses 1.409 electrons, resulting in a valence of +1.409, and provides a total of 2.818 electrons to the B-C cage. Four B atoms gain electrons, while twelve B atoms lose electrons. On average, each C atom gains 3.658 electrons, yielding a valence of -3.658.}

%Note the presence of both flat and dispersive bands near $E_F$, particularly the four-fold degenerate Dirac point at $\Gamma$ and the flat band near $M$, which are significant for potential superconductivity.    
\begin{figure}
\begin{center}
\includegraphics[width=8cm]{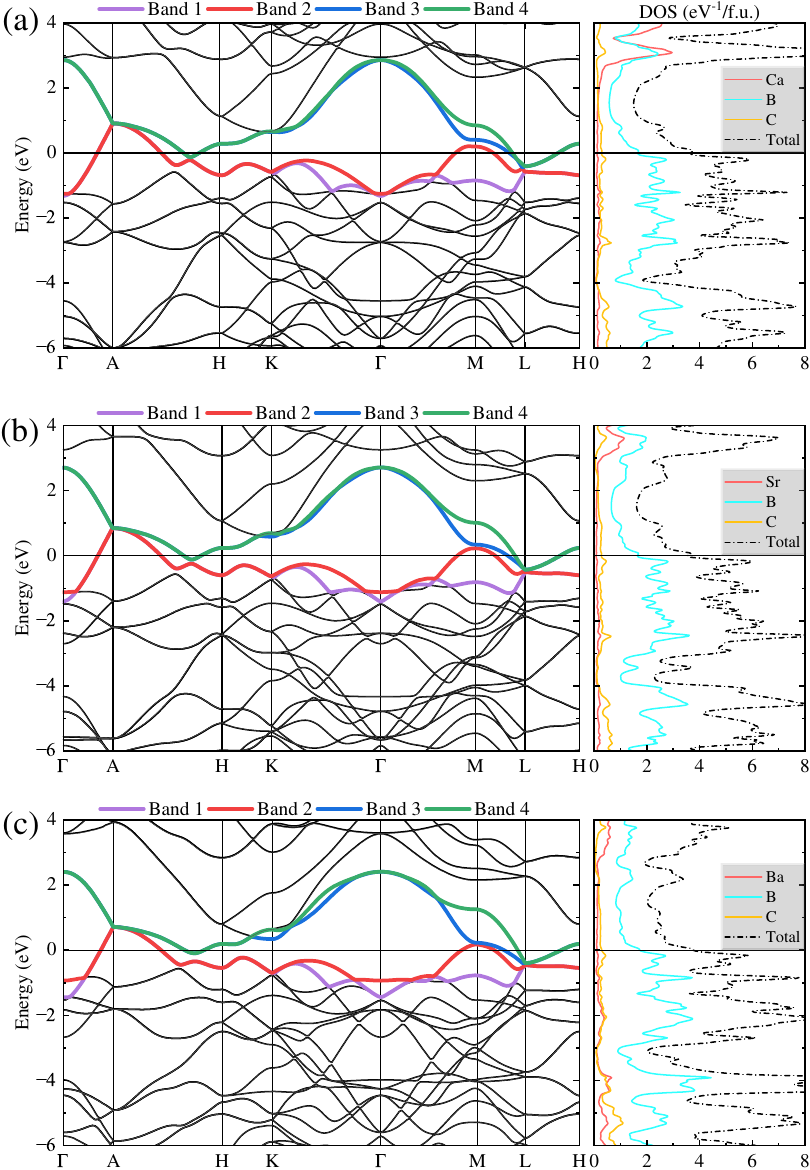}
\caption{
Electronic band structure and projected density of states (DOS) for the $P6_3/mmc$ structure at ambient pressure for (a) CaB$_8$C, (b) SrB$_8$C, and (c) BaB$_8$C. The band structures (left panels) are illustrated along high-symmetry paths in the Brillouin zone, highlighting four bands that cross the Fermi level ($E_F$). The corresponding DOS (right panels) reveal the contributions from individual atomic species, providing insight into the electronic properties of each compound. The $E_F$ is normalized to zero energy and is indicated by the solid horizontal line.
}
\label{fig3}
\end{center}
    \end{figure}
%\textcolor[rgb]{0.00,0.00,0.00}{To explore the potential superconductivity of the B-C framework, we employed the DFT method to calculate the properties of the X-B-C system. }
The electronic structure and physical properties of B-C inclusion complexes can be tailored by incorporating different guest atoms. We conducted a detailed investigation into the electronic band structure and density of states (DOS) for atoms in the boron-rich compound XB$_8$C, where $X$ represents Ca, Sr, and Ba. These compounds exhibit similar band structure characteristics, as shown in Fig.~\ref{fig3}. All compounds exhibit metallic characteristics, with four bands intersecting the Fermi level ($E_F$) along high-symmetry paths. These bands can be categorized into two distinct groups: (i) bands 1 and 2 show high degeneracy along the $\Gamma$-$A$-$H$-$K$ and $L$-$H$ paths, and (ii) bands 3 and 4 similarly exhibit degeneracy along these paths. \textcolor[rgb]{0.00,0.00,0.00}{Notably, we observed asymmetric band dispersions along the $\Gamma$-$A$-$H$ path: while the bands show linear behavior when approaching from the $\Gamma$ direction ($\Gamma$-$A$), they deviate from linearity when approaching from the H direction  ($H$-$A$).} The observed band structure configuration exhibits a compelling interplay of flat and dispersive bands near the Fermi level ($E_F$), a characteristic often associated with enhanced superconductivity \cite{simon1997superconductivity}. Particularly noteworthy is the presence of a flat band near $E_F$ along the $K-\Gamma-M$ and $L-H$ paths, coexisting with steep bands that cross $E_F$ along $\Gamma-A$. This unique combination of band features is especially significant, as it can lead to an increased density of states at $E_F$ and potentially stronger electron-phonon coupling. The flat band may contribute to a higher effective mass of charge carriers, while the steep bands can facilitate efficient charge transport. Such a band structure is conducive to the formation of Cooper pairs and could potentially result in a higher superconducting transition temperature. Analysis of the DOS reveals that the DOS for C and alkaline earth metal atoms remains relatively low. Conversely, B atoms contribute significantly more to the DOS at the Fermi level, indicating that the metallicity of these compounds primarily originates from the B atoms. This comprehensive analysis of the electronic structure provides crucial insights into the potential superconducting behavior of these boron-rich compounds and the role of different atomic species in determining their electronic properties.

%at the $\Gamma$ point, we observe a four-fold degenerate Dirac point approximately 1 eV above $E_F$,This feature is particularly significant as it suggests the presence of massless Dirac fermions, which could contribute to unique electronic properties and potentially influence the superconducting behavior of the material.

\begin{figure}
\begin{center}
\includegraphics[width=9cm]{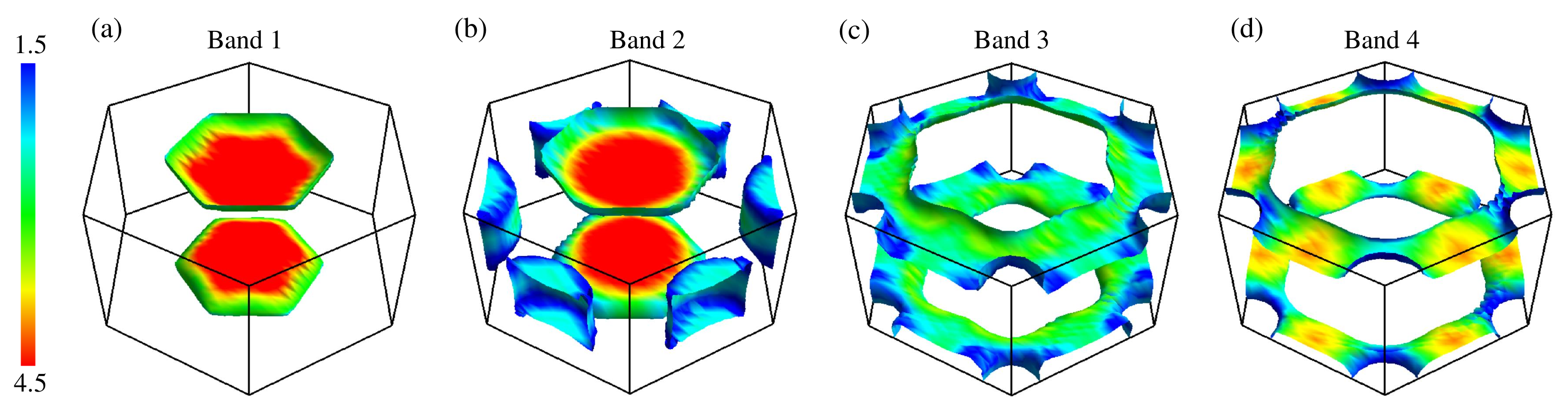}
\caption{%(a–d) Four separate parts constituting the entire $P6_3/mmc$-CaB$_8$C Fermi surface at 0 GPa, decorated with the Fermi velocity.
Fermi surface topology of $P6_3/mmc$-CaB$_8$C at ambient pressure. (a--d) Four distinct sheets of the Fermi surface, corresponding to the bands crossing the Fermi level. The color scale represents the magnitude of the Fermi velocity $|\mathbf{v}_F|$, with red indicating high velocities and blue indicating low velocities.
}
\label{fig4}
\end{center}
    \end{figure}

A detailed visualization of the Fermi surfaces for CaB$_8$C at ambient pressure is presented in Figure ~\ref{fig4}. The surfaces are color-coded to illustrate the distribution of Fermi velocity, with a gradient transitioning from blue to red as the velocity increases. This intricate Fermi surface topology comprises four distinct components: (i) Two pan-like surfaces located around the $A$ point at the top and bottom center of the BZ, exhibiting strong Fermi velocity at their centers; (ii) A similar structure to band 1, but featuring six additional bowl-shaped sheets located at the side faces of the hexagonal BZ; (iii) Flat, belt-like sheets extending across the top and bottom of the BZ; (iv) A structure akin to band 3, albeit characterized by different Fermi velocity distributions. Together, these components illustrate the complex electronic behavior of CaB$_8$C, highlighting the significance of Fermi surface topology in understanding its electronic properties. %The complex multi-sheet structure, featuring both large open surfaces and smaller closed pockets, which is indicative of the compound's three-dimensional electronic nature. 
    
\begin{figure}[!ht]
\begin{center}
\includegraphics[width=8cm]{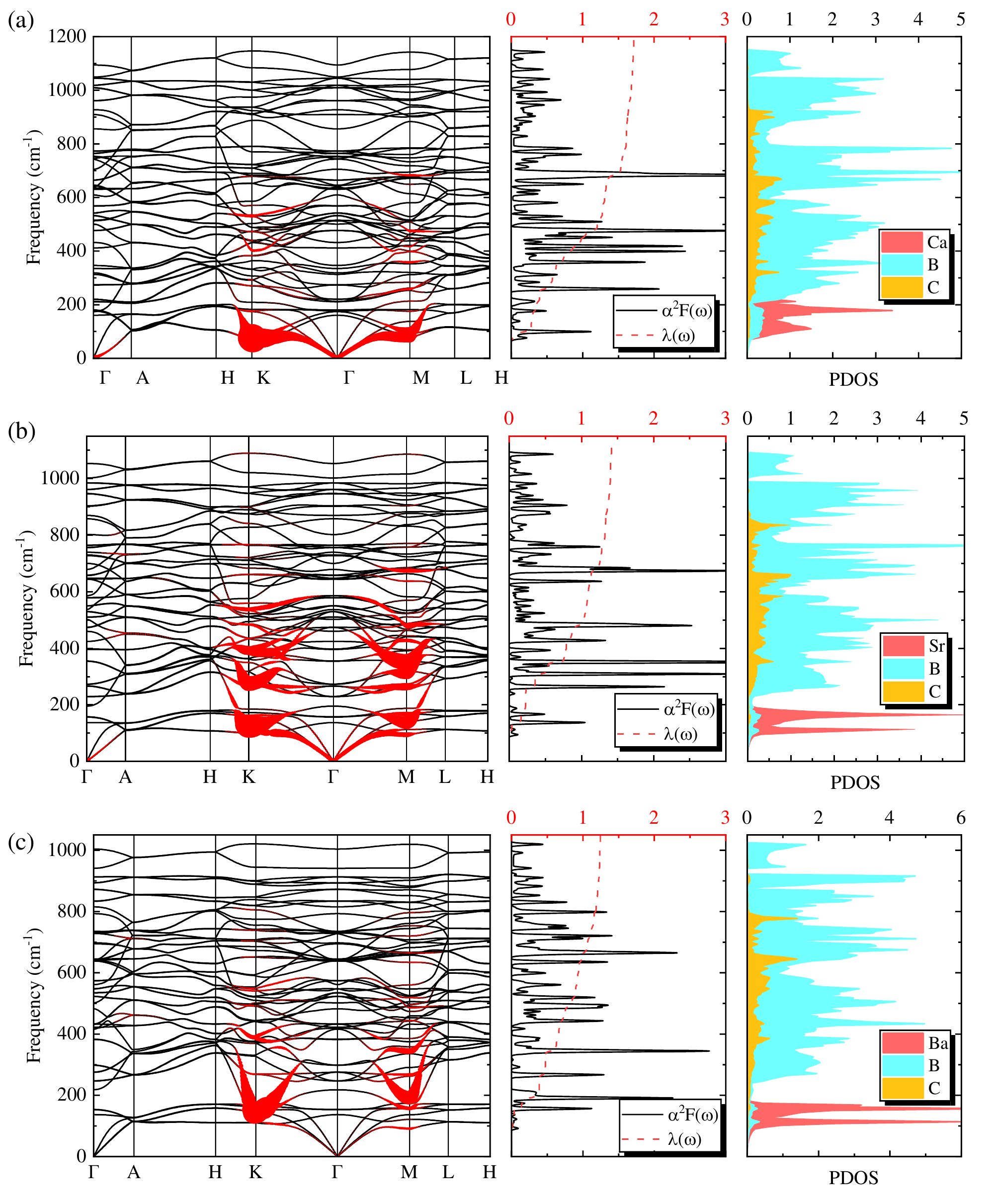}
\caption{
Phonon dispersions at ambient pressure for (a) CaB$_8$C, (b) SrB$_8$C, and (c) BaB$_8$C. The phonon frequencies $\omega_{qv}$ are depicted as black solid lines, while the color scale indicates the electron-phonon coupling strength $\lambda_{qv}$, represented by red circles. The middle panels present the Eliashberg spectral function $\alpha^{2}F(\omega)$ and the electron-phonon coupling strength $\lambda(\omega)$. The right panel displays the phonon density of states (PDOS) projected onto specific atomic species. In this diagram, contributions from Ca, Sr, and Ba are illustrated in red, while those from B and C are shown in cyan and yellow, respectively, facilitating clear differentiation of the atomic contributions to the phonon spectrum.
}
\label{fig5}
\end{center}
\end{figure}

\begin{table*}[!ht]
    \centering
    \begin{tabular}{cccccc}
     \hline \hline 
    ~ & $\lambda$ & $\omega$$_l$$_o$$_g$ (K) & $\overline{\omega_2}$ (K) & ADM $T_{c}$ (K) & Eliashberg $T_{c}$ (K)\\
    \hline
        CaB$_8$C & 1.715 & 490 & 702 & 77.1 & 83.9  \\ \hline
        SrB$_8$C & 1.416 & 522 & 687 & 64.4 & 68.9  \\ \hline
        BaB$_8$C & 1.244 & 506 & 681 & 53.2 & 55.8  \\ \hline
        RbBC$_8$ & 0.546 & 831 & 1008 & 14.4 & 15.4  \\ \hline
        NaB$_6$C$_3$ & 0.512 & 694 & 893 & 9.5 & 10.6  \\ \hline
        BaB$_7$C$_2$ & 0.628 & 590 & 741 & 15.9 & 17.1  \\ \hline
        SrB$_7$C$_2$ & 0.617 & 620 & 788 & 16.0 & 16.9  \\ \hline
    \hline
    \end{tabular}
     \caption{
     Several compounds exhibiting $T_{c}$ exceeding 10 K are compared based on their electron-phonon coupling strength $\lambda$, the logarithmic average of phonon frequency $\omega_{\text{log}}$, the root-mean-square frequency $\overline{\omega_2}$, and the calculated $T_{c}$. The $T_{c}$ values are obtained from both the Allen-Dynes modified McMillan  equation and the Eliashberg equation, with a Coulomb pseudopotential of $\mu^* = 0.1$.
\label{table1}}
\end{table*}

% \textcolor[rgb]{0.00,0.00,0.00}{Our investigation identified 24 dynamically stable compounds within the XB$_8$C, XB$_7$C$_2$, XB$_6$C$_3$, and XBC$_8$ structures (see Supplementary Material for phonon spectra)\cite{SupplementalMaterial}.} , as demonstrated in Figures S1-S18
 Further analysis focused on the phonon and electron-phonon coupling interactions of CaB$_8$C, SrB$_8$C, and BaB$_8$C. The calculations for XB$_8$C yield 40 phonon \textcolor[rgb]{0.00,0.00,0.00}{branches}, classified at the $\Gamma$ point according to the irreducible representation $D_{6h} = 3A_{1g} \oplus A_{1u} \oplus A_{2g} \oplus 5A_{2u}\oplus 5B_{1g} \oplus B_{1u} \oplus B_{2g} \oplus 3B_{2u} \oplus 4E_{2u} \oplus 6E_{2g} \oplus 6E_{1u} \oplus 4E_{1g}$, where $\Gamma_{\text{acoustic}} = A_{2u} \oplus E_{1u}$. As shown in Fig.~\ref{fig5}(a), the phonon dispersion at 0 GPa exhibits no imaginary frequencies, confirming dynamic stability. From Ca to Sr to Ba, we observe decreasing lattice vibrations and reduced maximum frequency in the phonon spectrum. The low-frequency branch ($<$200 cm$^{-1}$) is predominantly influenced by $X$ atom vibrations, while the high-frequency optical branch ($>$200 cm$^{-1}$) relates to B and C atom vibrations. Phonon dispersion calculations reveal strong electron-phonon coupling for CaB$_8$C along the $K$-$\Gamma$-$M$ path, particularly near 100 cm$^{-1}$. SrB$_8$C shows significant coupling at both $K$ and $M$ points across 0--500 cm$^{-1}$, whereas BaB$_8$C's coupling is mainly in the optical branches. The Eliashberg function $\alpha^2F(\omega)$ shows alkaline earth atom vibrations contributing to the acoustic mode and low-frequency density peak, while B and C vibrations affect the high-frequency optical branches. The detailed superconducting properties of several stable compounds are listed in Table \ref{table1}. Superconducting $\lambda$ values for CaB$_8$C, SrB$_8$C, and BaB$_8$C (1.715, 1.416, and 1.244, respectively) are primarily determined by high-frequency modes, decreasing with increasing atomic number of alkaline earth metals. The logarithmic average phonon frequency $\omega_{\text{log}}$ of CaB$_8$C is 490 K. Using a Coulomb pseudopotential $\mu^*$ of 0.1, the $T_c$ is estimated at 77.1 K (Allen-Dynes modified McMillan equation) and 83.9 K (Eliashberg equation). SrB$_8$C and BaB$_8$C exhibit slightly lower $T_c$ due to low-frequency vibrations of heavier alkaline earth metal atoms. \textcolor[rgb]{0.00,0.00,0.00}{We present the superconducting properties of additional X-B-C compounds (RbBC$_8$, NaB$_6$C$_3$, BaB$_7$C$_2$, and SrB$_7$C$_2$), as summarized in Table \ref{table1}. These materials exhibit moderate electron-phonon coupling constants ranging from 0.51 to 0.63, resulting in predicted superconducting transition temperatures between 10 and 20 K, suggesting their potential as moderate-temperature superconductors within the X-B-C family.} Our findings on SrB$_7$C$_2$ align with a recent study by Zhang et al. \cite{ZHANG2025113419}, further corroborating the potential of boron-carbon based compounds in high-temperature superconductivity research. Furthermore, the enrichment of boron increases the likelihood that the compounds will remain kinetically stable at ambient pressure while also enhancing the potential for higher superconducting transition temperatures. Compared to $\omega_{log}$, the $\lambda$ value has a more significant impact on $T_c$. For example, RbBC$_8$, a compound with a high $\omega_{log}$ of 831 K, has a $T_c$ of about 15 K, much lower than that of the XB$_8$C compounds. Consequently, three compounds with the highest boron enrichment: CaB$_8$C, SrB$_8$C, and BaB$_8$C have drawn our attention due to their higher $\lambda$ values and the correspondingly high calculated $T_c$ values. \textcolor[rgb]{0.00,0.00,0.00}{The phonon spectra for all other investigated compositions, encompassing both stable and unstable compounds, along with the additional anharmonic effects on the phonon spectra of XB$_8$C by combining the Stochastic Self-Consistent Harmonic Approximation (SSCHA) method\cite{monacelli2021stochastic, errea2013first, errea2014anharmonic, bianco2017second}, are detailed in the Supplementary Material\cite{SupplementalMaterial}. }

\section{Conclusion}

In this study, we employed first-principles methods to investigate the substitution of hydrides with hexagonal $P6_3/mmc$-structured B-C compounds, focusing on alkaline earth metal-boron-carbon inclusion complexes of the form XB$_8$C. Our comprehensive analysis encompassed stability, electronic structure, dynamic properties, and electron-phonon coupling interactions. While all examined compounds exhibit metastability under ambient conditions, CaB$_8$C notably becomes energetically favorable relative to its dissociated elements above 40 GPa, suggesting potential high-pressure synthesis routes. Electron-phonon coupling calculations reveal that XB$_8$C ($X$ = Ca, Sr, Ba) compounds maintain dynamic stability at 0 GPa and show promise as high-temperature superconductors, with CaB$_8$C predicted to have a superconducting transition temperature of 77.1 K at ambient pressure. These findings provide a robust foundation for further theoretical and experimental exploration of clathrate superconductors within the B-C framework, potentially opening new avenues for the design and synthesis of novel superconducting materials with enhanced properties and ambient condition stability. Future research directions may include investigating pressure-induced synthesis techniques, exploring doping strategies to further enhance $T_c$, and examining the potential of other metal-boron-carbon compositions for superconductivity.
%\clearpage
%---------------------------------------------------Acknowledgements-----------------------------------------------
\begin{acknowledgements}
%\ack{
This work is supported by the National Natural Science Foundation of China (Grants No. 12175107, 11504182), the Hua Li Talents Program of Nanjing University of Posts and Telecommunications, and the Natural Science Foundation of Nanjing University of Posts and Telecommunications (Grants No. NY224165, NY219087, NY220038).%\\}
\end{acknowledgements}
%\clearpage

%\bibliographystyle{unsrt}
%\bibliographystyle{apsrev4-2}
%\bibliography{bibfile}


\end{document}